\documentclass[superscriptaddress, amsmath, amssymb, aps, prx, showpacs, floatfix,10pt]{revtex4-2}

\usepackage{graphicx}
\usepackage{dcolumn}
\usepackage{bm}
\usepackage{dsfont}
\usepackage[colorlinks,linkcolor=blue,citecolor=blue,anchorcolor=blue]{hyperref}
\usepackage[mathlines]{lineno}
\usepackage{setspace}
\usepackage{xcolor}
\usepackage{amsmath,amssymb,booktabs,array}
\usepackage{textcomp} 
\usepackage{gensymb} 
\usepackage{soul} 
\usepackage{etoolbox} 
\usepackage{comment}
\usepackage{booktabs}    
\usepackage{array}       
\usepackage{dcolumn}  
\usepackage{rotating}
\usepackage{makecell}

\begin{document}


\title{\textbf{Analytical Theory of Higher-Order Collective Spin Interactions in Cavity Quantum Electrodynamics} 
}%

\newcommand{\bra}[1]{\left\langle {#1} \right|}
\newcommand{\ket}[1]{\left|  #1 \right\rangle}
\newcommand{\bracket}[3]{\langle {#1} | {#2} | {#3} \rangle}
\newcommand{\braket}[2]{\langle {#1} | {#2} \rangle}
\newcommand{\F}[0]{\ensuremath\mathcal{F}}
\newcommand{\aver}[1]{\ensuremath{\langle {#1} \rangle}}
\newcommand{\Nup}[0]{\ensuremath{N_\uparrow}}
\newcommand{\Ndown}[0]{\ensuremath{N_\downarrow}}
\newcommand{\var}[1]{\ensuremath{\left( \Delta #1 \right)^2}}

\definecolor{plotgreen}{RGB}{0,150,0}
\newcommand{\red}[1]{\textcolor{red}{#1}}

\newcommand{\scol}[1]{\textcolor{magenta}{#1}}

\author{Leilani Ainsworth}\thanks{These authors contributed equally to this work.}

\author{Chase Gomes}
\thanks{These authors contributed equally to this work.}

\author{Joseph Prescott}

\author{Kaley Wilcox}

\author{Jack Sullivan}

\author{Esteban Teran}

\author{Manav Bilakhia}

\author{Simone Colombo}%
 \email{Contact author: simone.colombo@uconn.edu}
\affiliation{%
Department of Physics, University of Connecticut, 196A Auditorium Road, Unit 3046, Storrs, Connecticut 06269-3046, USA.
}%

\date{\today}

\begin{abstract}
Cavity-mediated collective-spin interactions are commonly described by a quadratic one-axis twisting Hamiltonian. However, the underlying atom-light interaction naturally generates nonlinearities to arbitrary order. Here, we derive a closed-form analytical expression for the complete hierarchy of cavity-mediated collective-spin interactions. We show that the nonlinear coefficients $\chi_k$ are governed by Chebyshev polynomials, with $k$ the order of nonlinearity. This yields a universal scaling $\chi_k\propto\eta^k$ with the single-atom cooperativity $\eta$ and a description of their dependence on cavity detuning. The result provides a systematic framework for determining when higher-order nonlinearities become relevant and when the quadratic approximation breaks down. We identify experimentally relevant regimes in which higher-order terms substantially modify collective-spin dynamics, accelerating the generation of quantum correlations and quantum Fisher information, and demonstrate that finite-order expansions can accurately reproduce the full cavity-mediated evolution. Our results establish a general framework for understanding higher-order nonlinearities in cavity quantum electrodynamics and their role in collective entanglement and quantum-enhanced sensing.

\end{abstract}

\maketitle


\section{\label{sec:intro}Introduction}
The ability to engineer interactions among many quantum degrees of freedom lies at the heart of quantum simulation, quantum metrology, and quantum information science (QIS) \cite{Saf18,Kat22c,Luo24}. Among the available platforms, cavity quantum electrodynamics (QED) offers a distinctive route to collective dynamics, with confined photons mediating long-range and often all-to-all couplings between atoms \cite{Sre02,Sch09,Nor17, Li21}. This mechanism has enabled the deterministic generation of collective entanglement and metrologically useful many-body states including cavity spin squeezing \cite{Sch09,Ler11,Tor12,Lew18}, coherent cavity-mediated spin dynamics \cite{Nor17}, and interaction-based protocols for non-Gaussian quantum metrology \cite{Dav16,Col21, Li23}.

The theoretical understanding of these phenomena is based on an effective collective-spin description of cavity-mediated atom-light dynamics \cite{Li21} truncated at the quadratic order. These yield familiar pairwise nonlinear models such as one-axis twisting (OAT) and related Hamiltonians  \cite{Sch09,Bor17,Luo24}. Such a description has proven highly successful across a broad range of regimes. However, the underlying light-matter interaction naturally generates a hierarchy of higher-order terms whose role remains much less explored. Recent studies indicate that such terms can already become relevant in experimentally realistic cavity settings, where restricting the theory to second order may fail to capture important features of the full dynamics \cite{Gia25}. In parallel, cavity experiments have begun to isolate genuine multibody interactions by using optical dressing to suppress lower-order processes and reveal effective three- and four-body couplings~\cite{Luo25}.

These developments motivate a closer examination of cavity-mediated collective dynamics beyond the standard quadratic approximation. More generally, recent work on multibody collective interactions has shown that higher-order terms can enhance entanglement generation and metrological performance relative to conventional two-body twisting models \cite{Zha25}, while experiments in both trapped ions and cavities have demonstrated growing control over genuine multibody couplings \cite{Kat22,Kat22b,Kat22c,Luo25}. Together, these results suggest that higher-order collective nonlinearities should be viewed not only as corrections to the familiar quadratic theory but also as a useful framework for understanding collective dynamics in regimes where multibody effects become appreciable~\cite{Gia25,Luo25,Zha25}.

Here, we derive the complete arbitrary-order cavity-mediated nonlinear hierarchy in closed form, revealing a universal Chebyshev structure and the simple scaling $\chi_k\propto\eta^k$. We show that these higher-order terms can become dynamically important in experimentally accessible regimes, leading to qualitative departures from the conventional one-axis-twisting picture. We first establish the analytical structure and convergence of the expansion, then identify experimentally relevant regimes where higher-order terms become important. We finally quantify their impact on collective correlations, quantum Fisher information (QFI), and metrologically useful entanglement.

\begin{figure}[ht!]
    \centering
    \includegraphics[width=0.67\linewidth]{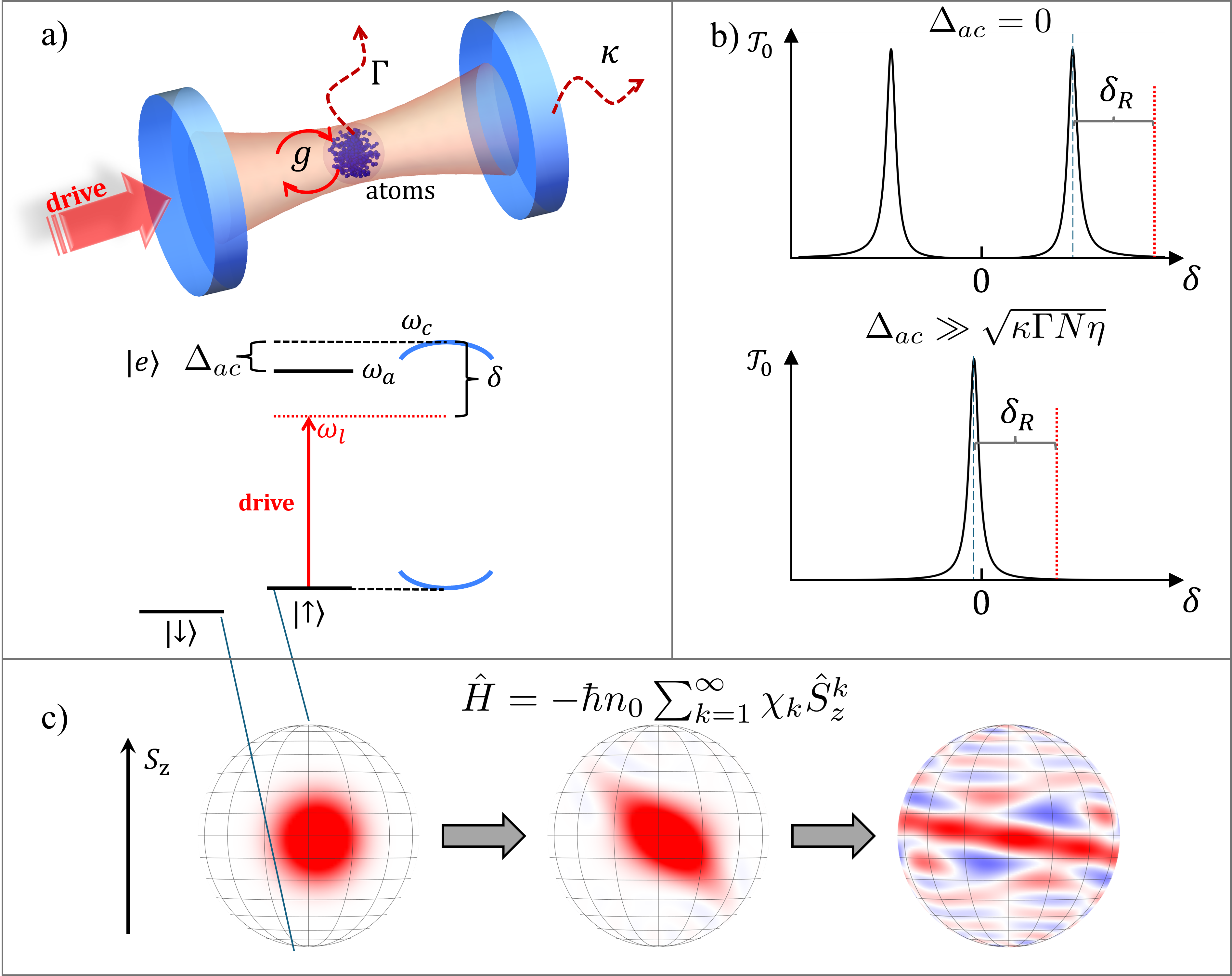}
    \caption{Cavity-mediated nonlinear dynamics of a collective atomic spin. \textbf{a)} Schematic of the atom--cavity system and relevant energy levels. An ensemble of $N$ three-level atoms is collectively coupled, through the transition $\left|   \uparrow\right\rangle\leftrightarrow\left|e\right\rangle$ and with coupling stregth $g$, to a single-mode cavity. The system is driven by a laser field with frequency $\omega_l$, with $\delta=\omega_l-\omega_c$ and $\Delta_{ac}=\omega_a-\omega_c$ denoting the laser--cavity and atom--cavity detunings, respectively. $\Gamma$ is the loss channel due to the incoherent scattering of a photon into free space by an atom, while $\kappa$ denotes the cavity loss. \textbf{b)} Spin-dependent cavity response arising in two distinct experimental regimes: resonant atom-cavity and dispersive cavity regimes. The collective spin projection $\hat S_z$ shifts the system response and, consequently, the intracavity photon number, providing the nonlinear feedback that generates the effective collective-spin Hamiltonian. \textbf{c)} Evolution of an initially coherent collective spin state (here, $N=20$), under the effective cavity-mediated Hamiltonian. The nonlinear spin-dependent interaction shears and distorts the collective-state distribution on the Bloch sphere, illustrating the generation of collective correlations beyond the conventional quadratic one-axis-twisting description} 
    \label{fig:setup}
\end{figure}
\section{\label{sec:CompleteHier} The Complete Hierarchy of Cavity-Mediated Nonlinear Spin Interactions}

We consider an ensemble of $N$ identical two-level atoms collectively coupled to a single-mode optical cavity and driven by a classical laser field \cite{Tan11,Li21,Sch09}. The drive is detuned from both the bare cavity resonance and the atomic transition. It is convenient to work with the dimensionless detunings
$$
 x_c = \frac{2\delta}{\kappa},
 \qquad
 x_a = \frac{2(\delta-\Delta_{ac})}{\Gamma},
$$
where $\delta = \omega_l - \omega_c$ is the laser--cavity detuning, $\Delta_{ac} = \omega_a - \omega_c$ is the atom--cavity detuning, and $\kappa$ and $\Gamma$ respectively denote the cavity and atomic linewidths. The strength of the light--matter interaction is set by the single-atom cooperativity, $\eta = 4g^2/(\kappa \Gamma)$, which compares the coherent atom--cavity coupling rate $g$ to the dominant dissipative channels~\cite{Sre02,Tan11,Li21}.
The atomic ensemble is described by collective spin operators $\hat{\mathbf S}=(\hat S_x,\hat S_y,\hat S_z)$, with $\hat S_z = \frac{1}{2}\sum_{i=1}^N \sigma_z^{(i)}$ representing the population imbalance between the two atomic states $\left|
\uparrow\right\rangle$ and $\left|
\downarrow\right\rangle$ \cite{Li21,Ma10}. 
The atoms shift the system's resonance by an amount that depends on $\hat S_z$, so that the intracavity intensity acquires a nonlinear dependence on the collective spin projection \cite{Mor07,Sch09,Ler11,Li21}.
This spin-dependent backaction is the origin of the effective collective interactions discussed below.
After eliminating the cavity field in the steady-state limit, the atom--light interaction gives rise to the effective spin Hamiltonian \cite{Sch09,Ler11,Li21,Cap23}

\begin{equation}
\hat H=-\hbar A(x_a)\,\hat n(\hat S_z)\,\hat S_z,
\label{eq:H_atomic}
\end{equation}

where $\hat n(\hat S_z)$ is the intracavity photon number operator and
$A(x_a)=\eta\mathcal L_d(x_a)/2$ is the dispersive light-shift coefficient. Here $\mathcal L_a(x)=1/(x^2+1)$ and $\mathcal L_d(x)=-x/(x^2+1)$ denote the absorptive and dispersive Lorentzian response functions. \cite{Tan11,Li21}. The dependence of $\hat n(\hat S_z)$ on the collective spin projection reflects the nonlinear backaction of the atomic ensemble on the cavity field and, consequently, of the cavity field back onto the atoms.

The intracavity photon number operator \cite{Li21, supmat} is

\begin{equation}
\label{eq:n_def}
\hat n(\hat S_z)=
\Omega
\left[
\left(
1+\left(\frac{N}{2}+\hat S_z\right)\eta\mathcal L_a(x_a)
\right)^2
+
\left(
x_c+\left(\frac{N}{2}+\hat S_z\right)\eta\mathcal L_d(x_a)
\right)^2
\right]^{-1},\nonumber
\end{equation}

where $\Omega$ is the driving strength. Defining the intracavity photon flux at $\hat S_z=0$,
\begin{equation}
n_0=
\Omega
\left[
\left(
1+\frac{N\eta}{2}\mathcal L_a(x_a)
\right)^2
+
\left(
x_c+\frac{N\eta}{2}\mathcal L_d(x_a)
\right)^2
\right]^{-1}\nonumber
\end{equation}
the Hamiltonian can be rewritten as

\begin{equation}
\hat H
=
-\hbar A(x_a)n_0
\frac{\hat n(\hat S_z)}{n_0}
\hat S_z.\label{eq:H_atomic_n0}
\end{equation}

This form separates the overall interaction scale determined by the average intracavity photon flux $n_0$ from the nonlinear dependence on the collective spin encoded in the dimensionless ratio $\hat n(\hat S_z)/n_0$. 
This normalization is particularly convenient because it naturally introduces the average number of photons entering the cavity as the relevant dynamical parameter. Since $n_0$ is the average photon flux through the cavity, the quantity
$n_{\mathrm{c}} = n_0 t$
is the average total number of photons that have interacted with the atomic ensemble after an evolution time $t$. Expressing the dynamics in terms of $n_{\mathrm{c}}$ provides a dimensionless interaction time that allows direct comparison between systems operating with different cavity linewidths and driving strengths.

With the definitions $\mathcal D=1-x_ax_c+\frac{N\eta}{2}$ and $\mathcal T_0\equiv n_0/\Omega$, where $\mathcal{T}_0$ is the collective cavity transmission at the mean-field operating point, the normalized photon number operator becomes

\begin{equation}
\frac{\hat n(\hat S_z)}{n_0}
=
\frac{1}
{1
+
2\eta\mathcal T_0\mathcal L_a(x_a) \mathcal D\,\hat S_z
+
\eta^2\mathcal T_0\mathcal L_a(x_a)\hat S_z^2}.\nonumber
\end{equation}

As derived in the Supplemental Material~\cite{supmat}, the latter can be expanded in terms of Chebyshev polynomials of the second kind $U_n(x)$ as 
\begin{equation}
\label{eq:n_expansion}
\frac{\hat n(\hat S_z)}{n_0}
=
\sum_{n=0}^{\infty}
\left(
\eta\sqrt{\mathcal T_0\mathcal L_a(x_a)}
\right)^n
U_n(x)\,
\hat S_z^n,\nonumber
\end{equation}
where we define $x= -\mathcal D\sqrt{\mathcal T_0\mathcal L_a(x_a)}$ for clarity.
This expansion in higher-order non-linear spin terms is valid for the Dicke states $\left|S, S_z\right\rangle$ satisfying 
\begin{equation}
\label{eq:convergence}
\left|S_z\right|<\frac{1}{\eta\sqrt{\mathcal T_0\mathcal L_a(x_a)}}.
\end{equation}
Physically, this bound is set by the extension of the many-body input state in $\hat S_z$.

Finally, substituting Eq.~\eqref{eq:n_expansion} into Eq.~\eqref{eq:H_atomic_n0} gives
\begin{equation}
\hat H
=
-\hbar n_0
\sum_{k=1}^{\infty}
\chi_k
\hat S_z^k,\label{eq:expansion}
\end{equation}

where the nonlinear interaction strengths are obtained in closed form as

\begin{equation}
\label{eq:chi_closed}
\chi_k=
\frac{\eta^k}{2}
\mathcal L_d(x_a)
\left(
\mathcal T_0\mathcal L_a(x_a)
\right)^{\frac{k-1}{2}}
U_{k-1}
\!\left(
-\mathcal D\sqrt{\mathcal T_0\mathcal L_a(x_a)}
\right).
\end{equation}

Equation~\eqref{eq:expansion} analytically decomposes the cavity-mediated dynamics into a hierarchy of well-defined nonlinear interaction orders, each with a quantitatively determined dependence on the experimentally relevant parameters. The convergence condition in Eq.~\eqref{eq:convergence} determines the range of validity of the expansion. The first few interaction strengths obtained from Eq.~\eqref{eq:chi_closed} are explicitly written in the supplemental material \cite{supmat}, and the first two terms $\chi_1$ and $\chi_2$ are the familiar linear light shift and quadratic shearing terms\cite{Sch09,Ler11,Bor17,Li21,Gia25}.

The powers of $\hat S_z$ can be interpreted as effective many-body interactions. Writing $\hat S_z^k=2^{-k}\sum_{i_1,\ldots,i_k}\sigma_z^{(i_1)}\cdots\sigma_z^{(i_k)}$ shows that each term contains contributions with both distinct and repeated atomic indices. For $k\ll N$, the distinct-index contributions dominate, so that $\hat S_z^k$ effectively describes a $k$-body interaction. As $k$ approaches $N$, repeated-index contributions become increasingly important, and for $k>N$ a genuine $k$-body interaction among distinct atoms is no longer possible. Nevertheless, $\hat S_z^k$ remains a well-defined collective operator, and Eq.~\eqref{eq:expansion} continues to define the hierarchy to arbitrary order.

This decomposition provides direct insight into the physical origin and relative importance of higher-order collective interactions, allowing one to determine when they can be neglected and when they become an essential part of the dynamics. More broadly, as multi-body interactions continue to emerge as a valuable resource in QIS and many-body physics, the present analytical framework provides the foundation for understanding and ultimately utilizing them in cavity QED systems.

A particularly interesting feature of this structure is the scaling with single-atom cooperativity,
\begin{equation}
\chi_k \propto \eta^k,
\end{equation}
which reflects the fact that each additional power of $\hat S_z$ arises from an additional order of cavity-mediated backaction. This result can also be understood in the picture introduced in Luo \textit{et al.}~\cite{Luo24, Luo25}, where a $k$-th order term in the collective spin Hamiltonian corresponds to a $2k$ virtual-photons process. Indeed, $\eta\propto g^2 $ indicates a two-photon process: one photon is absorbed by the atomic ensemble, and one photon is emitted into the cavity mode.

\begin{figure}[h!]
    \centering
    \includegraphics[width=0.85\linewidth]{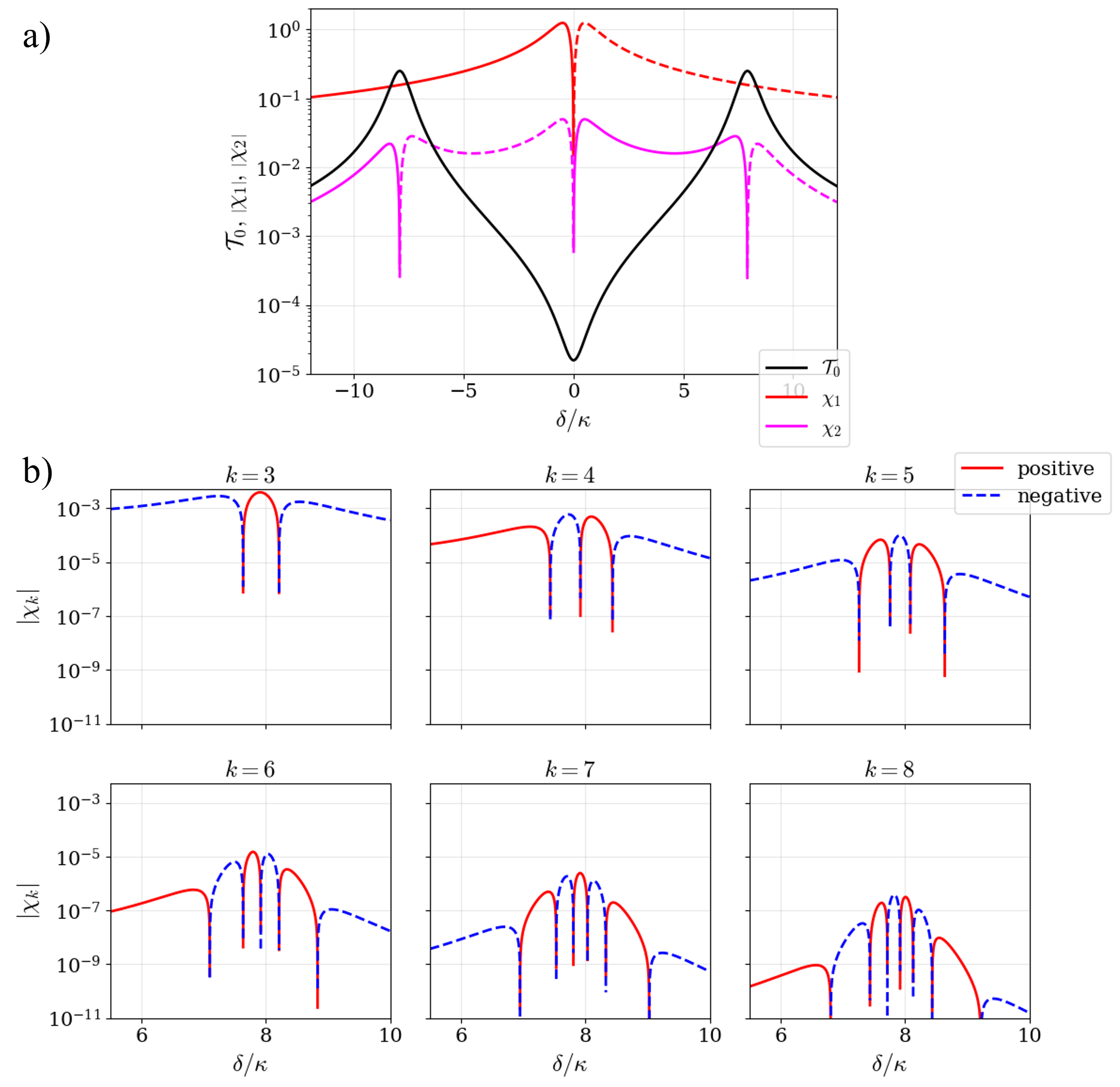}
    \caption{Higher-order nonlinear interaction strengths $\chi_k$ as a function of the drive detuning in the atom--cavity resonant regime  $\Delta_{ac}=0$. For visualization, the parameters are chosen as $N=100$, $\kappa=\Gamma=1$, and $\eta=5$. The progressively oscillatory structure with increasing order reflects the Chebyshev-polynomial hierarchy $\chi_k\propto U_{k-1}(x)$. Since $U_{k-1}(x)$ has $k-1$ zeros within $|x|<1$, the detuning dependence of the $k$th-order interaction contains $k-1$ nodes separating $k$ distinct lobes. Higher-order nonlinearities exhibit increasingly structured detuning dependence, with specific detunings at which individual interaction orders are suppressed or enhanced.}
    \label{fig:chikDetuning}
\end{figure}

Equation~\eqref{eq:chi_closed} also reveals the qualitative dependence of the nonlinear interaction strengths on the system parameters. Since the coefficients are proportional to the Chebyshev polynomials of the second kind, $\chi_k\propto U_{k-1}(x)$, their variation with cavity detuning inherits the oscillatory structure of these polynomials. In particular, the polynomial $U_{k-1}(x)$ possesses exactly ${k-1}$ simple zeros which translate into ${k-1}$ nodes of the interaction strength $\chi_k$. These nodes partition the detuning dependence into $k$ distinct lobes as shown in Fig.~\ref{fig:chikDetuning}. 
Consequently, increasing the interaction order produces progressively richer spectral structure and introduces additional detuning values at which a given nonlinear interaction is locally maximized or completely suppressed.   
The structure of the nonlinear interaction strengths provides a natural ordering principle for the expansion, but it does not by itself determine which terms are dynamically important. Their relevance is controlled jointly by the cooperativity, the detunings, and the range of collective-spin fluctuations sampled during the evolution. This point is especially important in view of recent work showing that higher-order corrections to cavity-mediated effective dynamics can already become relevant in experimentally realistic regimes and can be required to reproduce the behavior of the full atom--cavity evolution \cite{Gia25}. Quantifying the relative significance of these terms is therefore essential both for understanding their entangling power and for determining where the expansion can be truncated in practice. 

\section{Dynamical Signatures of Higher-Order Interactions}
\subsection{Spin-Correlation generation Rate}

To compare the importance of different orders in the effective Hamiltonian beyond their bare coupling strengths $\chi_k$, we look for a quantity that measures how fast they generate collective-spin correlations starting from an initially separable coherent spin state. The natural choice is the maximal transverse variance $\mathrm{Var}(\hat S_{\perp,\theta})$, with the collective spin generator $\hat S_{\perp,\theta}=\cos\theta\,\hat S_y+\sin\theta\,\hat S_z$. Physically, this quantity measures the growth of the principal axis of the spin-noise distribution. Since nonlinear collective interactions generate quantum correlations by deforming and rotating this distribution, the maximal transverse variance provides a natural measure of the strength of the induced many-body dynamics~\cite{Win94, Bra94, Kit93, Ma10, Riv07}. This is the notion of short-time entanglement growth for states initialized along the $x$ axis~\cite{Kit93,Jin09,Opa14b,Riv07}.

We consider an initial coherent spin state polarized along $x$, $|\psi_0\rangle = \sum_{S_z=-S}^{S} c_{S_z}\,|S,S_z\rangle$, with
$c_{S_z} \equiv \sqrt{\binom{2S}{S+S_z}}\Big(\tfrac12\Big)^S$,
and analyze its evolution under a single nonlinear term proportional to $\hat S_z^k$. In the large-$S$ limit, the short-time dynamics is captured by the transverse covariance matrix \cite{Mad04,Opa14b,Per19}
\begin{equation}
\tilde\Sigma_\perp(n_c, k)
=
\begin{pmatrix}
\sigma^2_y(n_c, k) & c(n_c, k) \\
c(n_c, k) & 1
\end{pmatrix},\nonumber
\end{equation}
where
$
\sigma^2_y \equiv \frac{2\,\mathrm{Var}(\hat S_y)}{S}$
and
$c \equiv \frac{2\,\mathrm{Cov}(\hat S_y,\hat S_z)}{S}$
are the variance and covariance normalized to the coherent-spin-state variance $S/2$. The lower-right entry remains equal to unity because $\hat S_z$ is conserved under $\hat S_z^k$ evolution; The nonlinear dynamics shear the uncertainty distribution within the transverse plane while leaving the $S_z$ fluctuations unchanged.
The maximal and minimal normalized transverse variances are then the eigenvalues of $\tilde\Sigma_\perp(n_c)$
\begin{equation}
\sigma^2_{\max,\min}
=
\frac{\sigma^2_y+1}{2}
\pm
\sqrt{\left(\frac{\sigma^2_y-1}{2}\right)^2+c^2},\nonumber
\end{equation}
which are the principal transverse variances of the covariance matrix \cite{Riv07}. 

At short times, one finds in the Gaussian approximation \cite{Opa14b,Per19, supmat}
\begin{equation}
\sigma^2_y(n_c, k)
\simeq
1 + 2k^2 S\bigl[\mu_{2k-2}-\mu_{k-1}^2\bigr](\chi_k n_c)^2,
\qquad
c(n_c, k)
\simeq
-2k\,\mu_k\,\chi_k n_c \nonumber
\end{equation}
where $\mu_n=\langle \hat S_z^n\rangle_0$ are the moments in the initial coherent spin state~\cite{supmat}. Therefore, for all odd orders $k$ the covariance is 0. The covariance $c(n_c)$ appears at linear order in time ($n_c$), reflecting an initial tilt of the uncertainty, while the growth of $\sigma_y^2(n_c)$ is quadratic in time, describing the broadening of fluctuations along a fixed axis. These formulas recover the standard OAT variances for $k=2$ \cite{Kit93,Jin09,Li21} and extend naturally to higher-order nonlinearities.

Finally, to quantify the dynamical importance of the different nonlinear interactions, we consider the initial growth rate of the transverse spin fluctuations,
\begin{equation}
R_k\equiv\frac{\partial^2}{\partial n_c^2} \sigma_y^2(n_c, k) = 8k^2\left(\frac{S}{2}\right)^k\chi_k^2
\begin{cases}
(2k-3)!!, & k \text{ even},\\[1mm]
(2k-3)!!-(k-2)!!^2, & k \text{ odd}.
\end{cases},\label{eq:generationSpeed}
\end{equation}
which measures how rapidly the collective-spin uncertainty departs from that of an initially coherent spin state. Since
$\sigma_y^2=1$ for a coherent spin state, this quantity directly characterizes the rate at which the nonlinear dynamics generate collective-spin correlations. This makes it the relevant quantity for comparing the physical importance of different nonlinear orders.

The dynamics only probes Dicke states within the range of $S_z$ populated by the initial state. We take as a practical convergence bound for the expansion the point at which the criterion in Eq.~\eqref{eq:convergence} is first violated for $|S_z| = \sqrt{N}$, corresponding to twice the standard deviation of the initial coherent spin state, $\sqrt{\mathrm{Var}(\hat S_z)} = \sqrt{N}/2$. This choice reflects that the evolved state is predominantly supported on Dicke states within this range, so that the expansion remains valid over the bulk of the relevant Hilbert space even though formally Eq.~\eqref{eq:convergence} must hold for every populated $|S, S_z\rangle$ Dicke state.

\begin{figure} [ht!]
    \centering
    \includegraphics[width=0.95\linewidth]{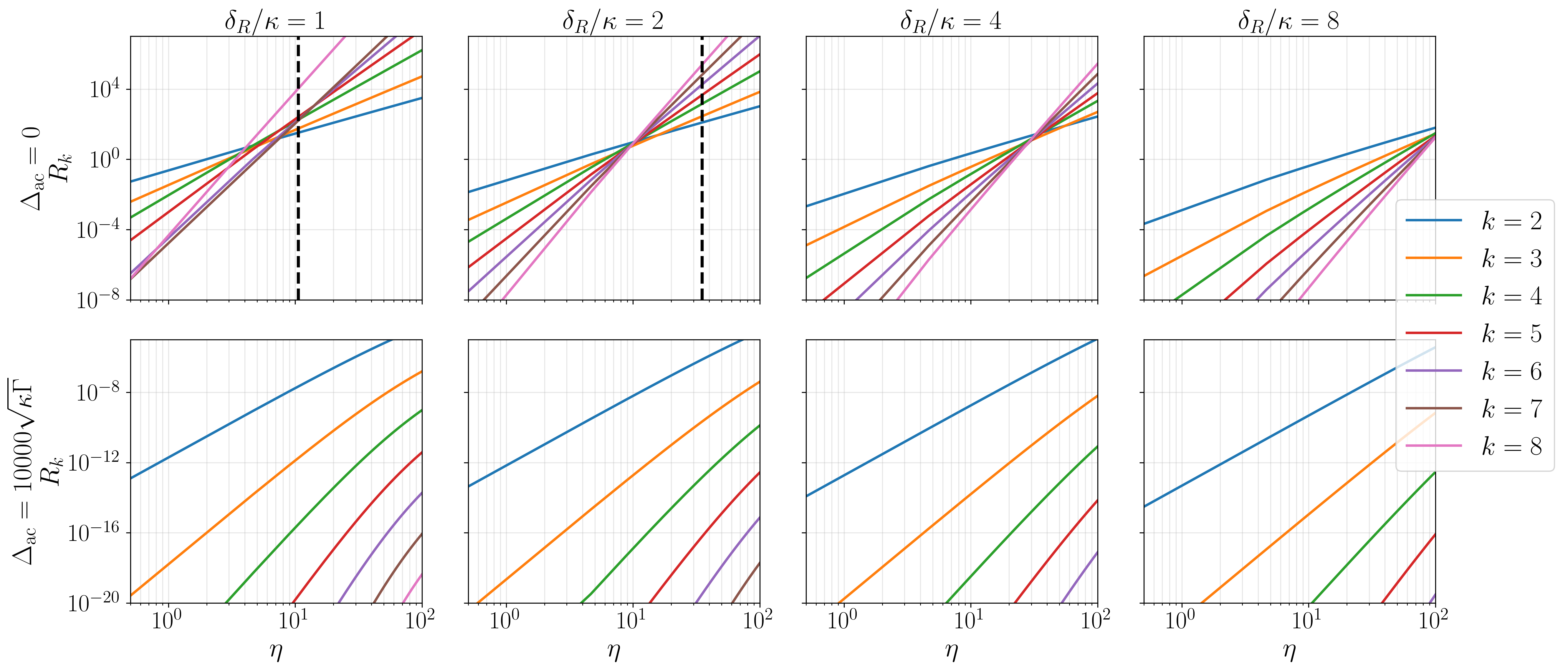}
    \caption{Higher-order contributions to the generation of transverse spin correlations. We set $\kappa=\Gamma$ and $N=1000$. The plotted quantity $R_k$ characterizes the rate at which the $k$th-order nonlinear term generates transverse spin correlations. The upper row corresponds to operation near atom-cavity resonance ($\Delta_{ac}=0$), while the lower row shows the far-detuned dispersive regime ($\Delta_{ac}\gg\sqrt{\kappa\Gamma N\eta}$). Each column corresponds to a different drive detuning $\delta_R$ relative to the dressed cavity resonance. The dashed lines indicate the boundary of validity of the expansion given by Eq.~\eqref{eq:convergence}; for larger $\eta$, the expansion no longer converges over the relevant spin-state support. The results illustrate that higher-order nonlinearities can contribute significantly to the generation of collective correlations as the single-atom cooperativity is increased, consistent with the analytical scaling $\chi_k\propto\eta^k$ of the nonlinear coefficients.}
    \label{fig:sigma_y}
\end{figure}

Figure~\ref{fig:sigma_y} compares the correlation-generation rate $R_k$ for nonlinear interactions up to eighth order as a function of the single-atom cooperativity. We fix the atom number $N$ and the drive detuning relative to the dressed cavity resonance $\delta_R$. The $\mathcal{T}_0$ remains therefore independent of $\eta$. In the atom-cavity resonant case, $\Delta_{ac}=0$, the dressed-cavity resonance scales as $\sqrt{\eta}$ and, consequently, also $\delta_R$ and $x_a\sim\sqrt{\eta}$. Therefore, $\mathcal{L}_a\sim\eta^{-1}$ and $\mathcal{L}_d\sim\eta^{-1/2}$. Consequently, (\ref{eq:chi_closed}) gives $\chi_k\propto\eta^{k/2}$ and therefore $R_k\propto\eta^k$.  In the far-detuned regime, $\Delta_{ac}\gg\sqrt{\kappa\Gamma N\eta}$, the atomic detuning $x_a$ remains approximately constant as $\eta$ is varied, such that $\mathcal{L}_a$ and $\mathcal{L}_d$ are independent of $\eta$. In this case, $\chi_k\propto\eta^k$ and $R_k\propto\eta^{2k}$. Thus, as clearly visible in Fig.~\ref{fig:sigma_y}, the scaling of the correlation-generation rate differs qualitatively between the two regimes. Moreover, we notice that in the atom-cavity resonant regime, the contributions of higher-order terms become significant already at moderate $\eta$. In the cavity-detuned regime, despite the faster growth of $R_k$ with $\eta$, higher-order terms remain comparatively suppressed due to the relative magnitude of their prefactors. We discuss these prefactors and their role in the relative contribution of the higher-order terms in the following subsection.

\subsection{Comparison with Existing Experiments}

As derived in the supplemental material~\cite{supmat}, the coefficients $\chi_k$ obey a rigorous bound that follows directly from $0<\mathcal T_0\leq1$ and $|U_{k-1}(x)|\leq k$,
\begin{equation}
|\chi_k| \;\leq\; \frac{k\,\eta^k}{2}\,|\mathcal L_d(x_a)|\,\mathcal L_a(x_a)^{\frac{k-1}{2}}.
\end{equation}
In the dispersive cavity regime, $\Delta_{ac}\gg\sqrt{\kappa\Gamma N\eta}$, the atom-cavity detuning satisfies $x_a\gg1$, and the Lorentzian responses reduce to $\mathcal L_a(x_a)\approx x_a^{-2}$, $\mathcal L_d(x_a)\approx -x_a^{-1}$, so that
\begin{equation}
|\chi_k| \;\lesssim\; \frac{k}{2}\,\frac{\eta^k}{x_a^{\,k}}.
\end{equation}
 Using the parameters reported by Davis \textit{et al.}~\cite{Dav20}, $|x_a| = 2|\Delta_{ac}|/\Gamma = 2\times10\,\mathrm{GHz}/6\,\mathrm{MHz}\approx3333$, $\eta=5.1$, and $N\approx100{,}000$, the exact coefficients of Eq.~\eqref{eq:chi_closed} give $R_2/R_3\approx28{,}571$: the quadratic term generates spin correlations more than four orders of magnitude faster than the third-order one.

The other regime is realized near the atom--cavity resonance, $\Delta_{ac} \approx 0$. In this regime the atom-cavity detuning $x_a$ no longer provides a large parameter on its own, and the nonlinear couplings $\chi_k$ can, in principle, become significant already at moderate cooperativities. Indeed, as shown in Fig.~\ref{fig:sigma_y}, in this regime higher-order interactions become significant at currently more accessible single-atom cooperativities. However, if the drive is detuned from the dressed cavity resonance by a $\delta_R$ much larger than the resonance width, one finds $|x|\approx 1$ and $\mathcal{T}_0\approx{x_c^{-2}}$, thereby adding a further suppression on top of the bound above. The relevant limit then becomes
\begin{equation}
\left|\chi_k \right| \approx \frac{k}{2}\frac{\eta^k}{x_a^k\,x_c^{k-1}}.
\end{equation}
Here, the large $\delta_R$ implies a significant additional suppression factor beyond the dispersive case. For representative parameters $\eta\approx8$ and $\delta_R/\kappa\sim10$, roughly matching the conditions of Colombo \textit{et al.}~\cite{Col21}, the quadratic term generates spin correlations about $100$ times faster than the third-order term.

Across both classes of experiments, higher-order nonlinearities have remained hidden behind the same quadratic OAT Hamiltonian~\cite{Ler09,hos16,Dav20,Gre22,Bra19,Col21}, with the exception of Luo \textit{et al.}~\cite{Luo25}, where optical dressing is used to cancel lower-order interactions and directly expose three- and four-body terms. In the Supplemental Material, we report the ratios $R_3/R_2$ and $R_4/R_2$ across a broader set of existing cavity QED experiments, confirming that the quadratic term dominates throughout. The absence of observed higher-order signatures in these cases reflects the specific detuned operating regimes explored, where detuning-dependent prefactors counter the intrinsic $\eta^k$ enhancement and keep the quadratic interaction dominant over the accessible timescales. This is not indicative of the fundamental irrelevance of higher-order terms. The suppression identified above accounts only for the coherent generation of spin correlations; non-unitary processes such as atomic and cavity dissipation act as a further hindrance, competing with higher-order coherent dynamics before they can accumulate observable signatures.

The parameter regime explored in this work is directly accessible to current and near-term cavity QED platforms. Ensembles of $N\sim1000$ atoms collectively coupled to an optical cavity with single-atom cooperativities in the range $\eta\sim1$~to~$100$ have already been realized experimentally~\cite{kawasaki2019geometrically, Haa14, Pet25}. These developments place several of the regimes identified above, where higher-order nonlinearities produce measurable deviations from the quadratic approximation, within reach of existing and near-future hardware. Current cavity QED platforms are therefore capable of operating in parameter windows where the suppressions discussed above are substantially reduced. In this regime, higher-order collective interactions are expected to accelerate the generation of collective spin correlations and many-body entanglement, leading to experimentally observable deviations from the conventional OAT dynamics before decoherence limits the evolution.

\section{Accuracy of the Truncated Hierarchy}
We now assess the accuracy of the expansion \eqref{eq:expansion} and the relative coefficients \eqref{eq:chi_closed} at the level of the quantum state. We compare the state generated by the complete effective Hamiltonian with that obtained from finite-order truncations. Throughout this analysis, we consider $N=1000$ atoms initialized in a coherent spin state on the equator of the generalized Bloch sphere and quantify the deviation using the state infidelity
\begin{equation}
\mathcal{I}=1-\left|\langle \psi_{\rm full}(n_c)|\psi_{\rm trunc}(n_c)\rangle\right|^2,
\end{equation}
where $|\psi_{\rm full}(n_c)\rangle$ denotes the state evolved under the complete effective Hamiltonian and $|\psi_{\rm trunc}(n_c)\rangle$ the corresponding state obtained by truncating the expansion. We compare two truncations: the second-order term, corresponding to the conventional OAT Hamiltonian, and a representative eighth-order truncation.

\begin{figure}[t!]
    \centering
    \includegraphics[width=0.95\linewidth]{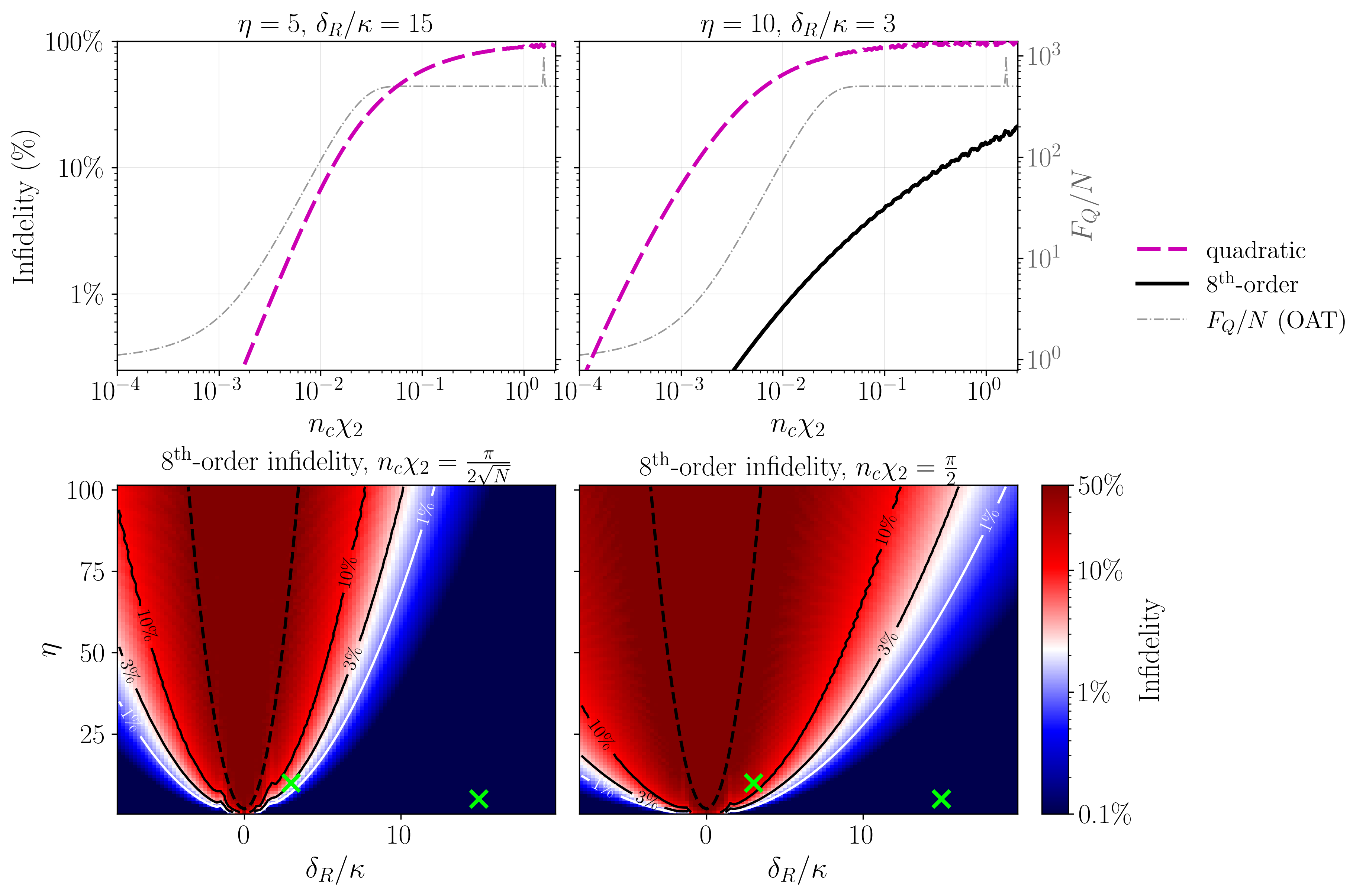}
\caption{
Validation of the closed-form Hamiltonian expansion against the full cavity-mediated dynamics, for $N=1000$ atoms initialized in a coherent spin state on the equator of the generalized Bloch sphere. Without loss of generality, we chose here $\Gamma=\kappa$.
\textbf{Top} Infidelity between the full Hamiltonian and its truncation at second order (quadratic, purple dashed) or eighth order (black solid), as a function of the reduced evolution time $n_c\chi_2$, for $(\eta,\delta_R/\kappa)=(5,13)$ (left) and $(10,3)$ (right). The eighth-order expansion remains accurate over several decades in $n_c\chi_2$ where the quadratic truncation has already failed. The normalized QFI $F_Q/N$ of the exact evolution (gray dashed-dotted, right axis) is shown alongside for reference, illustrating that substantial infidelity growth in the quadratic approximation occurs already before $F_Q/N$ approaches its plateau.
\textbf{Bottom} Infidelity of the eighth-order truncation relative to the full Hamiltonian, as a function of cooperativity $\eta$ and Rabi detuning $\delta_R/\kappa$, evaluated at the near-Heisenberg squeezing time $n_c\chi_2=\pi/(2\sqrt N)$ (left) and the OAT GHZ-generation time $n_c\chi_2=\pi/2$ (right). White and black solid contours mark 1\%, 3\%, and 10\% infidelity; the black dashed curve marks the convergence boundary of Eq.~\eqref{eq:convergence}, defined by the extent of the initial state in $\hat S_z$. The eighth-order expansion remains accurate to within a few percent over most of the formally convergent region.
The green crosses denote the combinations of parameters $eta$ and $\delta_R/\kappa$ used in the top line.}
\label{fig:benchmarking}
\end{figure}

We first examine how the truncation error accumulates during the evolution. The top row of Fig.~\ref{fig:benchmarking} shows the infidelity as a function of the reduced evolution time $\chi_2 n_c$ for two representative parameter regimes. The quadratic truncation initially follows the full dynamics but its infidelity grows rapidly with evolution time, even at the more detuned operating point $(\eta,\delta_R/\kappa)=(5,13)$. At the more nonlinear point $(\eta,\delta_R/\kappa)=(10,3)$, the breakdown is correspondingly more pronounced. In contrast, the eighth-order truncation remains accurate over several decades in evolution time. The rapid growth of the quadratic-truncation error therefore occurs well before the full dynamics reaches the most strongly entangled regime, illustrating that on experimentally relevant evolution times, the validity of the OAT approximation can be lost. 

To quantify the accuracy of the eighth-order expansion more systematically, the bottom row of Fig.~\ref{fig:benchmarking} shows its infidelity relative to the full Hamiltonian across the $(\eta,\delta_R/\kappa)$ parameter space at two evolution times motivated by quantum metrology. The shorter time, $\chi_2 n_c=\pi/(2\sqrt{N})$, corresponds to the near-Heisenberg spin-squeezing regime of OAT, where the metrological gain has already approached within approximately $3\,\mathrm{dB}$ of the Heisenberg limit \cite{Kit93, Monz11}. The longer time, $\chi_2 n_c=\pi/2$, corresponds to the generation of a maximally entangled Greenberger-Horne-Zeilinger (GHZ) state under ideal OAT dynamics. The eighth-order expansion remains accurate to within a few percent over most of the formally convergent region at both evolution times, demonstrating that a finite-order truncation can faithfully reproduce the full cavity-mediated dynamics well beyond the regime in which the quadratic approximation remains valid.

\subsection{Metrological implications}

In cavity-mediated spin squeezing, the metrological gain is conventionally understood as a reduction of quantum projection noise below the standard quantum limit, quantified by the Wineland parameter,
$\xi^2~=~\frac{N\,\mathrm{Var}(\hat S_{\perp,\min})}{|\langle\hat{S}\rangle|^2}$~\cite{Win94},
which measures the sensitivity gain in Ramsey interferometry relative to an unentangled ensemble. A more general measure of metrologically useful entanglement is the QFI, which for pure states and generator $\hat S_{\perp,\theta}$ is $F_Q = 4\,\mathrm{Var}(\hat S_{\perp,\theta})$~\cite{Bra94, Pez07, Ma10, pezze2018quantum}; maximizing over $\theta$ gives $F_Q/N = \sigma_{\max}^2$, so the optimized transverse variance introduced above is directly the QFI per particle. For Gaussian states, the two quantities coincide and $\xi^{-2}$ saturates the full available metrological advantage.

Conventional treatments of cavity-mediated entanglement generation truncate the collective-spin Hamiltonian at second order (OAT), for which the state remains approximately Gaussian until the curvature of the generalized Bloch sphere becomes appreciable. Only at this comparatively late stage does $\xi^{-2}$ depart from $F_Q/N$; Over most of the relevant evolution, the Wineland parameter alone suffices to characterize the state's metrological utility.
\begin{figure}[ht!]
    \centering
    \includegraphics[width=0.905\linewidth]{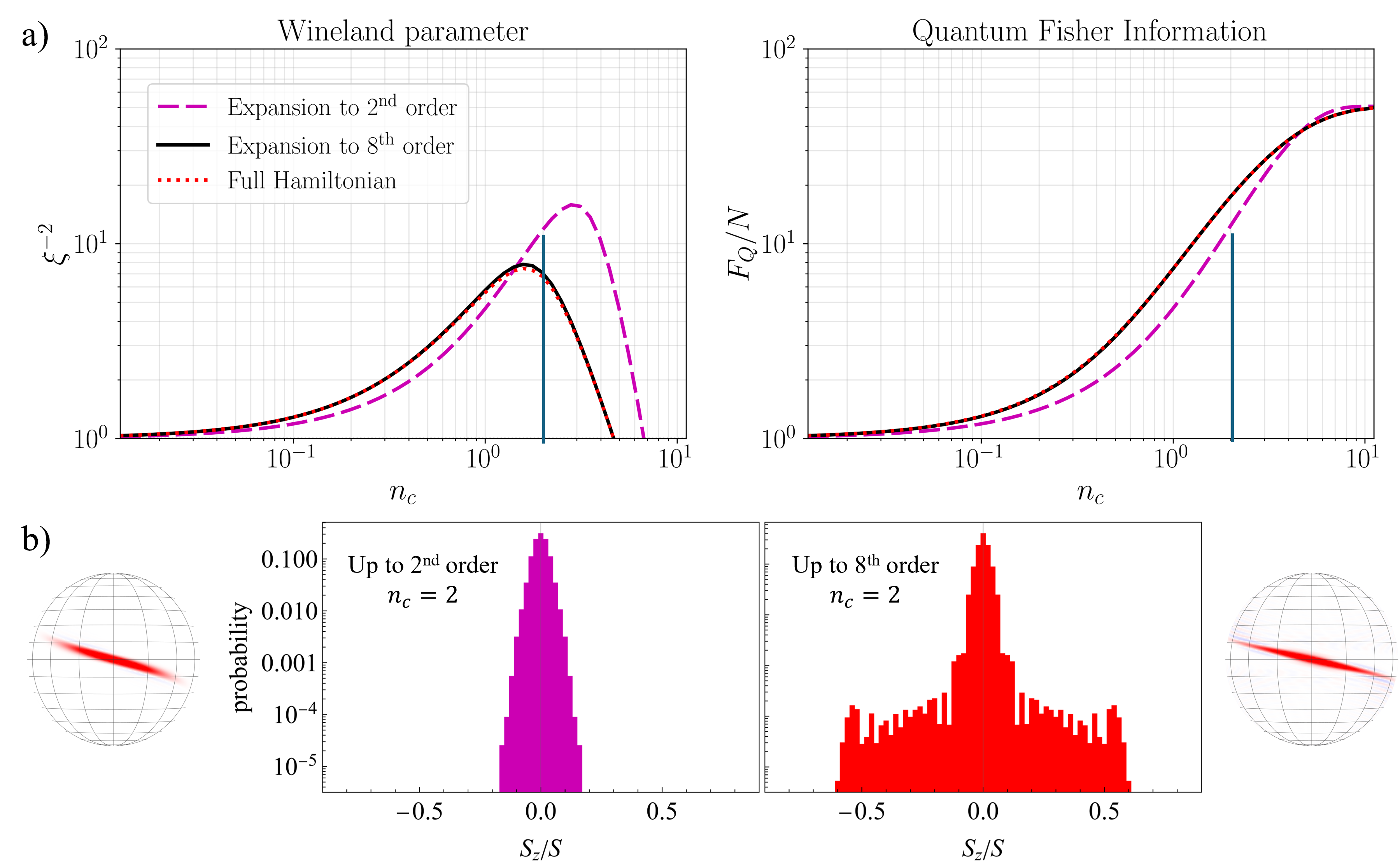}
    \caption{Comparison of the metrological performance predicted by the quadratic one-axis-twisting Hamiltonian (gray) and the complete cavity-mediated Hamiltonian (red) for $N=100$ atoms, $\eta=20$, $\Delta_{ac}=0$, and $\delta_R=4\sqrt{\kappa\Gamma}$. Only even-order nonlinear terms are retained; odd-order contributions, including the linear light shift, are removed by a spin-echo sequence. \textbf{a)} Inverse Wineland parameter $\xi^{-2}$ (left) and QFI per particle $F_Q/N$ (right) vs.\ average intracavity photon number $n_c$. The dotted curve (eighth-order truncation) is indistinguishable from the full Hamiltonian. The blue line marks $n_c=2$, where the states in b) are evaluated. \textbf{b)} Probability distribution along the optimal quadrature and corresponding Wigner distribution for the quadratic (left) and full (right) Hamiltonian.}    \label{fig:WinelandParamComparison}
\end{figure}

As we show in Fig.~\ref{fig:WinelandParamComparison}, retaining the higher-order terms changes this picture qualitatively: the state departs from Gaussianity far earlier than Bloch-sphere curvature alone would predict, and $\xi^{-2}$ and $F_Q/N$ diverge already within the near-resonant, experimentally accessible regime. Compared to the OAT Hamiltonian, the Wineland parameter degrades significantly earlier under the complete Hamiltonian evolution (Fig.~\ref{fig:WinelandParamComparison}(a), left). The additional nonlinearities distort the spin distribution and generate pronounced non-Gaussian tails (Fig.~\ref{fig:WinelandParamComparison}(b)) that inflate the minimum quadrature variance entering $\xi^2$. Despite the presence of stronger underlying correlations influenced by higher-order terms, the Wineland parameter degrades.

However, the growth of QFI (see Fig.~\ref{fig:WinelandParamComparison}(a), right) shows that the correct accounting of the higher-order terms in the Hamiltonian generates entanglement at least as fast as, and over most of the evolution substantially faster than, the OAT Hamiltonian. The earlier degradation of $\xi^{-2}$ is, therefore, due to the failure of Gaussian (linear Ramsey) readout to access the metrological resources encoded in an increasingly non-Gaussian state, and not to lower quantum correlations. The point at which $\xi^{-2}$ and $F_Q$ diverge marks the onset of a regime where linear readout is no longer optimal and where more advanced protocols, like interaction-based or Bayesian estimation protocols~\cite{Str14, Nol17, Baa21, Col21,Li23}, are required to recover the full sensitivity stored in the non-Gaussian state. 

More broadly, this result illustrates how the closed-form expression \eqref{eq:chi_closed} predicts directly, from a platform's cooperativity and detuning, where the Gaussian approximation will break down, and standard readout will cease to be optimal. This is without resorting to full numerical simulation. Because this threshold varies strongly across parameter space, such a prediction is inherently system-specific, making the analytical hierarchy derived here a crucial design tool for extracting maximal metrological performance from any cavity-mediated entangling platform.

\section{Conclusion}
We have developed a general analytical framework for cavity-mediated collective-spin interactions beyond the conventional quadratic approximation. By exploiting the generating function of Chebyshev polynomials of the second kind, we derived a closed-form expression for the complete hierarchy of nonlinear interaction strengths to arbitrary order together with an explicit convergence criterion for the effective Hamiltonian. This analytical solution replaces the successive evaluation of higher-order derivatives with a compact universal expression that directly determines every nonlinear coefficient from the system parameters.

The analytical structure of the solution reveals several general properties of cavity-mediated nonlinear interactions. In particular, we showed that the $k$th-order interaction strength scales as the $k$th power of the single-atom cooperativity, while its dependence on cavity detuning follows directly from the corresponding Chebyshev polynomial. These results establish a simple organizing principle for higher-order cavity-mediated interactions and provide physical insight into how their magnitude, sign, and spectral structure emerge from the underlying atom-cavity dynamics.

Using this framework, we identified experimentally relevant parameter regimes where higher-order nonlinearities substantially modify the collective dynamics. We demonstrated that they accelerate the generation of collective correlations and QFI relative to the conventional one-axis-twisting description, determined where the quadratic approximation ceases to be quantitatively reliable, and benchmarked successive higher-order truncations against the full cavity-mediated Hamiltonian. Together, these results clarify when higher-order interactions must be retained to accurately describe cavity-mediated many-body dynamics.

Beyond the specific system considered here, the analytical approach developed in this work provides a general tool for studying collective interactions in cavity QED.
The expansion is not restricted to states near the equator, and extending it to states with larger $\langle \hat S_z\rangle_0$ or broader support could make the framework applicable to the study of higher-order interaction dynamics in lower-cooperativity multi-body platforms such as in Luo \textit{et al.}~\cite{Luo25}.
It enables rapid evaluation of arbitrary-order effective Hamiltonians as well as offers a transparent connection between emergent many-body interactions and cavity parameters at the individual atom level. It also establishes a foundation for exploring interaction engineering, non-Gaussian state generation, and quantum-enhanced sensing in regimes where, rather than being a perturbative correction, higher-order nonlinearities become a valuable resource.
We anticipate that this framework will facilitate the systematic design and control of cavity-mediated many-body interactions, opening new opportunities for quantum simulation, entanglement generation, and precision metrology beyond the conventional one-axis-twisting paradigm.

\begin{acknowledgments}
We thank Vladan Vuleti{\'c}, Zeyang Li, Edwin Pedrozo-Pe{\~n}afiel, and James K. Thompson for valuable discussions. We acknowledge support from ONR and University of Connecticut startup funds.
\end{acknowledgments}

\bibliographystyle{apsrev4-2}
\bibliography{apssamp}


\clearpage
\onecolumngrid

\begin{center}
\textbf{\large Supplemental Material: Analytical Theory of Higher-Order Collective Spin Interactions in Cavity Quantum Electrodynamic}

\vspace{0.3cm}

Leilani Ainsworth, Chase Gomes, Joseph Prescott, Kaley Wilcox, Jack Sullivan, Esteban Teran, Manav Bilakhia, and Simone Colombo

\vspace{0.2cm}

\textit{Department of Physics, University of Connecticut, 196A Auditorium Road, Unit 3046, Storrs, Connecticut 06269-3046, USA}
\end{center}

\setcounter{section}{0}
\renewcommand{\thesection}{S\arabic{section}}
\renewcommand{\thefigure}{S\arabic{figure}}
\renewcommand{\thetable}{S\arabic{table}}
\renewcommand{\theequation}{S\arabic{equation}}

\section{Quantum Mechanical Derivation of the intracavity photon-number operator}
\label{sec:SM_intracavity_field}

Here, we derive the intracavity photon-number operator used in Eq.~\ref{eq:n_def} of the main text with a quantum mechanical treatment, starting from the master equation. The result agrees with the one obtained with a semiclassical approach in Tanji-Suzuki \textit{et al.}~\cite{Tan11} and Li \textit{et al.}~\cite{Li21}
We consider a single cavity mode coupled uniformly to an ensemble of $N_\uparrow$ identical two-level atoms. The cavity mode has frequency $\omega_c$, the atomic transition frequency is $\omega_a$, and the atom--cavity coupling strength is $g$. The cavity and atomic energy decay rates are denoted by $\kappa$ and $\Gamma$, respectively.

The dynamics of the driven atom--cavity system are described by the master equation
\begin{equation}
\dot{\rho}
=
-\frac{i}{\hbar}
\left[
\hat{H}_{\rm lab},
\rho
\right]
+
\kappa \mathcal{L}[\hat{a}]\rho
+
\Gamma
\sum_{j=1}^{N_\uparrow}
\mathcal{L}[\hat{\sigma}_{j}^{-}]\rho ,
\label{eq:SM_master}\nonumber
\end{equation}
where
$\mathcal{L}[\hat{O}]\rho
=
\hat{O}\rho\hat{O}^{\dagger}
-\frac{1}{2}
\left\{
\hat{O}^{\dagger}\hat{O},
\rho
\right\}$, is the Lindblad quasi-operator and
\begin{equation}
\frac{\hat{H}_{\rm lab}}{\hbar}
=
\omega_c \hat{a}^{\dagger}\hat{a}
+
\omega_a
\sum_{j=1}^{N_\uparrow}
\hat{\sigma}_{j}^{+}\hat{\sigma}_{j}^{-}
+
g
\sum_{j=1}^{N_\uparrow}
\left(
\hat{a}^{\dagger}\hat{\sigma}_{j}^{-}
+
\hat{a}\hat{\sigma}_{j}^{+}
\right)
+
\Omega
\left(
\hat{a}^{\dagger}e^{-i\omega_l t}
+
\hat{a}e^{i\omega_l t}
\right).
\label{eq:SM_Hlab}\nonumber
\end{equation}
Here, $\omega_l$ is the frequency of the coherent driving field and $\Omega$ is the effective cavity driving strength, including the coupling between the external field and the cavity.

We consider here the weak-drive regime, in which the system remains predominantly in the collective ground state. in this condition, the Lindbladians reduce to trivial quantum jumps, and we obtain the non-Hermitian Hamiltonian,
\begin{equation}
\hat{H}_{\rm NH}
=
\hat{H}_{\rm lab}
-
i\hbar\frac{\kappa}{2}\hat{a}^{\dagger}\hat{a}
-
i\hbar\frac{\Gamma}{2}
\sum_{j=1}^{N_\uparrow}
\hat{\sigma}_{j}^{+}\hat{\sigma}_{j}^{-}.
\label{eq:SM_HNH_lab}\nonumber
\end{equation}
The imaginary terms account for cavity and atomic decay. The corresponding quantum-jump processes are contained in the full master equation, Eq.~\eqref{eq:SM_master}. In the weak-drive, linear-response regime, the coherent amplitudes can therefore be obtained from $\hat{H}_{\rm NH}$.

Transforming to a frame rotating at the drive frequency $\omega_l$ gives
\begin{equation}
\frac{\hat{H}_{\rm NH}}{\hbar}
=
\left(
-\delta-i\frac{\kappa}{2}
\right)
\hat{a}^{\dagger}\hat{a}
+
\left(
\Delta_{ac}-\delta-i\frac{\Gamma}{2}
\right)
\sum_{j=1}^{N}
\hat{\sigma}_{j}^{+}\hat{\sigma}_{j}^{-}
+
g
\sum_{j=1}^{N}
\left(
\hat{a}^{\dagger}\hat{\sigma}_{j}^{-}
+
\hat{a}\hat{\sigma}_{j}^{+}
\right)
+
\Omega
\left(
\hat{a}^{\dagger}+\hat{a}
\right),
\label{eq:SM_Hrot}\nonumber
\end{equation}
where we defined the cavity--drive and atom--cavity detunings $\delta \equiv \omega_l-\omega_c$, and
$\Delta_{ac}\equiv\omega_a-\omega_c$.

To first order in the drive strength, only the zero- and one-excitation manifolds are required. We denote the collective atomic ground state by $\lvert G\rangle$ and introduce the symmetric single-excitation state
\begin{equation}
\lvert B\rangle
=
\frac{1}{\sqrt{N}}
\sum_{j=1}^{N}
\lvert e_j\rangle ,\nonumber
\end{equation}
where $\lvert e_j\rangle$ denotes the state in which atom $j$ is excited. Because all atoms couple identically to the cavity, only this symmetric bright state couples to the cavity field, with an enhanced coupling strength $\sqrt{N}g$.

In the one-excitation manifold spanned by
\begin{equation}
\left\{
\lvert B,0\rangle,
\lvert G,1\rangle
\right\},\nonumber
\end{equation}
the relevant part of the effective non-Hermitian Hamiltonian is
\begin{equation}
\frac{\hat{H}_{\rm NH}^{(1)}}{\hbar}
=
\begin{pmatrix}
\Delta_{ac}-\delta-i\Gamma/2
&
\sqrt{N}g
\\[4pt]
\sqrt{N}g
&
-\delta-i\kappa/2
\end{pmatrix}.
\label{eq:SM_H1}
\end{equation}

We treat the drive perturbatively and write the state as
\begin{equation}
|\psi\rangle
\simeq
|G,0\rangle
+
\Omega|\psi_1\rangle =|G,0\rangle
+
c_e|e,0\rangle
+
c_c|G,1\rangle ,\nonumber
\end{equation}
where $c_e$ and $c_c$ are the amplitudes of the atomic and cavity excitations, respectively.

The cavity-field amplitude is obtained directly by projecting the first-order state onto the one-photon cavity state,
\begin{equation}
c_c
\equiv
\langle G,1|\psi\rangle .\nonumber
\end{equation}
Using the one-excitation Hamiltonian~\eqref{eq:SM_H1}
, this gives
\begin{equation}
c_c
=
\frac{-2i\Omega}{\kappa}
\left[
1-i\frac{2\delta}{\kappa}
+
\frac{N_\uparrow\eta}
{1-i\frac{2(\Delta_{ac}-\delta)}{\Gamma}}
\right]^{-1} =
\frac{-2i\Omega}{\kappa}
\left[
1-i\,x_c
+
\frac{N_\uparrow\eta}
{1-i\,x_a}
\right]^{-1},
\label{eq:SM_cc_eta} \nonumber
\end{equation}
where we used the single-atom cooperativity
$
\eta
=
\frac{4g^2}{\kappa\Gamma}$. 

The corresponding intracavity photon number is therefore
\begin{equation}
n
=
|c_c|^2 =
\Omega
\left[
\left(
1+N_\uparrow\eta\mathcal L_a(x_a)
\right)^2
+
\left(
x_c+N_\uparrow\eta\mathcal L_d(x_a)
\right)^2
\right]^{-1},\nonumber
\end{equation}

The number of atoms participating in the cavity interaction is determined by the population in the $\lvert\uparrow\rangle$ state. We therefore promote $N_\uparrow$ to the corresponding population operator,
\begin{equation}
N_\uparrow
\rightarrow
\hat{N}_\uparrow
=
\hat{S}_z+\frac{N}{2},
\label{eq:SM_Nup_operator}\nonumber
\end{equation}
where $\hat{S}_z$ is the collective spin projection and $N$ is the total number of atoms. Since $\hat{S}_z$ is diagonal in the $N_{\uparrow}$ basis and the cavity response is smooth in $N_{\uparrow}$, one can perform this operator substitution in the weak-drive expansion.
The intracavity photon number consequently becomes an operator-valued function of the collective spin,
\begin{equation}
\hat{n}(\hat{S}_z)
=
\Omega
\left[
\left(
1+\left(\hat{S}_z+\frac{N}{2}\right)\eta\mathcal L_a(x_a)
\right)^2
+
\left(
x_c+\left(\hat{S}_z+\frac{N}{2}\right)\eta\mathcal L_d(x_a)
\right)^2
\right]^{-1}.
\label{eq:SM_n_operator}
\end{equation}

\section{Closed-form generating function and convergence condition}


Starting from the Gegenbauer generating function

\begin{align}
\frac{1}{(1-2xt+t^{2})^{\alpha}} = \sum_{n=0}^{\infty}C_{n}^{({\alpha})}(x)t^{n}\nonumber
\end{align}

In the case where $\alpha = 1$, we get the generating function for Chebyshev polynomials of the second kind $U_n(x)$~\cite{arfken1968mathematical}

\begin{align}
\frac{1}{1-2xt+t^{2}} = \sum_{n=0}^{\infty}U_{n}(x)t^{n}, \lvert x \rvert \leq 1, \lvert t \rvert<1\label{eq:SM_generatingfunction}
\end{align}

To obtain the closed-form generating function and convergence criteria, we match $\frac{\hat n(\hat S_z)}{n_0}$ to the generating function of \eqref{eq:SM_generatingfunction} here above

\begin{align}
    \frac{1}{1-2xt+t^{2}} = \frac{\hat n(\hat S_z)}{n_0} = \frac{1}{1+2\eta\mathcal{T}_0\mathcal{L}_{a}\mathcal{D}\hat{S_{z}} + \mathcal{T}_{0} \mathcal{L}_{a}\eta^{2}\hat{S_{z}^{2}}}\nonumber
\end{align}

It is clear that $t = \eta\sqrt{\mathcal{T}_{0}\mathcal{L}_{a}(x_a)}\hat{S_{z}} = C_{0}\hat{S_{z}}$
and
$x = -\mathcal{D}\sqrt{\mathcal{T}_{0}\mathcal{L}_{a}(x_a)}.$
Interestingly, $\left|x\right|\leq 1$ for all real parameters, satisfying automatically one of the condition for generating function of $U_n(x)$.
As a consequence of the convergence condition for the Chebyshev series, $\lvert t \rvert < 1$, we obtain convergence condition for the higher-order expansion
\begin{align}
    \lvert\hat{S}_{z}\rvert < \frac{1}{\eta \sqrt{\mathcal{T}_0\mathcal{L}_{a}(x_a)}}
\end{align}

The generating function becomes 
$$
\sum_{n=0}^{\infty}U_{n}(x)t^{n} = \sum_{n=0}^{\infty}C_{0}^{n}U_{n}(x)\hat{S_z^{n}}
$$
Therefore, we finally obtain the expansion

\begin{align}
   \frac{\hat{n}(\hat{S}_z)}{n_{0}} = \sum_{n=0}^{\infty}C_{0}^{n}U_{n}(x)\hat{S_z^{n}}\label{eq:supp_expansion}
\end{align}

Substitute \eqref{eq:supp_expansion} into \eqref{eq:H_atomic_n0}.

$$
    \hat{H} = -\hbar A(x_{a})n_{0}\sum_{n=0}^{\infty}C_{0}^{n}U_{n}(x)\hat{S_{z}}^{n+1}
$$

Let $k = n+1 \rightarrow n = k-1$

$$
    \hat{H} = -\hbar A(x_{a})n_{0}\sum_{k=1}^{\infty}C_{0}^{k-1}U_{k-1}(x)\hat S_z^k = -\hbar n_0 \sum_{k=1}^{\infty} \chi_k \, \hat S_z^k
$$

We get the closed-form expression for $\chi_k$

\begin{align}
    \chi_{k}& = A(x_{a})C_{0}^{k-1}U_{k-1}(x)\nonumber\\
    & =\frac{\eta^k}{2}\mathcal{L}_d(x_a) \left(\mathcal{T}_{0}\mathcal{L}_{a}(x_a)\right)^{\frac{k-1}{2}}U_{k-1}\left(-\mathcal{D}\sqrt{\mathcal{T}_{0}\mathcal{L}_{a}(x_a)}\right).\label{eq:supmat_Chi}
\end{align}

Here we explicitly report the results up to the 8-th order:

\begin{align}
    \chi_1 &= \tfrac{1}{2}\eta\,\mathcal{L}_d(x_a), \nonumber \\
    \chi_2 &= -\eta^2\,\mathcal{T}_0\,\mathcal{L}_a(x_a)\mathcal{L}_d(x_a)\,\mathcal{D}, \nonumber \\
    \chi_3 &= 2\eta^3\,\mathcal{T}_0\,\mathcal{L}_a(x_a)\mathcal{L}_d(x_a)\left(\mathcal{T}_0\,\mathcal{L}_a(x_a)\mathcal{D}^2-\tfrac{1}{4}\right),\nonumber \\
    \chi_4 &= 2\eta^4\mathcal{T}_0^2\mathcal{L}_a^2(x_a)\mathcal{L}_d(x_a)\,\mathcal{D}\left(1-2\mathcal{T}_0\mathcal{L}_a(x_a)\mathcal{D}^2\right),\nonumber \\
    \chi_5 &= \tfrac{1}{2}\eta^5\mathcal{T}_0^2\mathcal{L}_a^2(x_a)\mathcal{L}_d(x_a)\left(1+4\mathcal{T}_0\mathcal{L}_a(x_a)\mathcal{D}^2\left(4\mathcal{T}_0\mathcal{L}_a(x_a)\mathcal{D}^2-3\right)\right),\nonumber \\
    \chi_6 &= -\eta^6\mathcal{T}_0^3\mathcal{L}_a^3(x_a) \mathcal{L}_d(x_a)\,\mathcal{D}\left(4\mathcal{T}_0\mathcal{L}_a(x_a)\mathcal{D}^2-3\right)\left(4\mathcal{T}_0\mathcal{L}_a(x_a)\mathcal{D}^2-1\right), \nonumber \\
    \chi_7 &= \tfrac{1}{2}\eta^7\mathcal{T}_0^3\mathcal{L}_a^3(x_a)\mathcal{L}_d(x_a)\left[8\mathcal{D}^2\mathcal{T}_0\mathcal{L}_a(x_a)(2\mathcal{D}^2\mathcal{T}_0\mathcal{L}_a(x_a)-1)(4\mathcal{D}^2\mathcal{T}_0\mathcal{L}_a(x_a)-3)-1\right], \nonumber \\
    \chi_8 &= -4\eta^8\mathcal{T}_0^4\mathcal{L}_a^4(x_a)\mathcal{L}_d(x_a)\mathcal{D}(2\mathcal{D}^2\mathcal{T}_0\mathcal{L}_a(x_a)-1)\left[1+8\mathcal{D}^2\mathcal{T}_0\mathcal{L}_a(x_a)(8\mathcal{D}^2\mathcal{T}_0\mathcal{L}_a(x_a)-1)\right]. \nonumber
\end{align}

\section{Short time large-S limit}

Here we present the derivation of the transverse variance growth under a general polynomial twisting Hamiltonian in the large S expansion limit\cite{Opa14b, Per19}.

Consider a coherent spin state polarized along $x$ in the fully symmetric Hilbert subspace with total spin $S=N/2$. The initial moments are

\begin{align}
\langle \hat S_x\rangle &= S,
\nonumber \\
\langle \hat S_y\rangle &= \langle \hat S_z\rangle = 0,
\nonumber
\end{align}

and

\begin{align}
\mathrm{Var}(\hat S_y) &= \mathrm{Var}(\hat S_z) = \frac{S}{2}.
\nonumber
\end{align}

In the limit $S\gg1$, the distribution of $S_z$ is Gaussian to leading order, with variance $S/2$ and moments
\begin{align}
\mu_n \equiv \langle S_z^n\rangle
&\simeq
\begin{cases}
0, & n \text{ odd}, \\
(n-1)!!\,\left(\frac{S}{2}\right)^{n/2}, & n \text{ even}.
\end{cases}
\label{eq:app_gaussian_moments}
\end{align}
Moreover, the finite differences are approximated by derivatives
$$
(S_z+k)^n - S_z^n \simeq k\,n\,S_z^{n-1}, \qquad k=1,2.
$$

Take a general polynomial Hamiltonian

\begin{align}
\hat H &= f(\hat S_z)= \hbar n_0\sum_{k=2}^{\infty} \chi_k \hat S_z^k,
\label{eq:app_poly_ham}
\end{align}

with fixed, non-scaled couplings $\chi_k$. Since the Hamiltonian commutes with $\hat S_z$, $\hat S_z$  is a constant of motion (i.e., $\hat S_z(t)=\hat S_z$, exactly. In the Heisenberg picture, the evolution of the state is then given by 
\begin{align}
\frac{d \hat{S}_y}{dt}
&= \frac{i}{\hbar}[\hat H,\hat S_y] = -n_0\sum_{k=2}^{\infty} \chi_k \sum_{r=0}^{k-1} \hat S_z^r \hat S_x \hat S_z^{k-1-r}.
\label{eq:app_heisenberg_sy_exact}
\end{align}

At short times around the $x$ polarized coherent spin state, we replace $\hat S_x$ by its mean value $S$ at leading order \cite{Kit93, Opa14b}. This yields

\begin{align}
\hat S_y(t)
&= \hat S_y-St\,f'(\hat S_z)+O(t^2),
\nonumber
\end{align}

with

\begin{align}
f'(\hat S_z)
&=\hbar n_0\sum_{k=2}^{\infty} k\chi_k \hat S_z^{k-1}.
\nonumber
\end{align}

This relation is the starting point for the transverse second moments.

It is convenient to normalize all transverse variances to the coherent spin state variance $S/2$. Define

$$
\sigma^2_y(n_c)
=\frac{2\,\mathrm{Var}(\hat S_y)(n_c)}{S}, \quad 
\sigma^2_z(n_c) = \frac{2\,\mathrm{Var}(\hat S_z)(n_c)}{S}=1, \quad 
c(n_c) = \frac{2\,\mathrm{Cov}(\hat S_y,\hat S_z)(n_c)}{S},
$$
where $n_c\equiv n_0 t$ indicates the average number of photons that passed through the cavity. 
Using $\langle \hat S_y\rangle=\langle \hat  S_z\rangle=0$, the covariance is

\begin{align}
\mathrm{Cov}(\hat S_y,\hat S_z)(n_c)
&= \frac{1}{2}\langle\hat  S_y(n_c)\hat S_z+\hat S_z\hat S_y(n_c)\rangle = -S\,\langle \hat S_z f'(\hat S_z)\rangle\,n_c + O(n_c^2).
\nonumber
\end{align}

Therefore, the normalized covariance is

\begin{align}
c(n_c)
&\approx -2\frac{n_c}{n_0}\,\langle \hat S_z f'(\hat S_z)\rangle = -2n_c\sum_{k=2}^{\infty} k\chi_k\mu_k=-2n_c\sum_{\substack{k=2 \\ k\,\mathrm{even}}}^{\infty}
   k\chi_k (k-1)!!\left(\frac{S}{2}\right)^{k/2}
\label{eq:app_c_general}
\end{align}
for $k$~even, and vanishes for $k$~odd; In the Gaussian equatorial state, only even powers contribute at leading order.
 
For the variance of $S_y$, the same short time expansion gives

\begin{equation}
\mathrm{Var}(\hat S_y)(n_c)
= \mathrm{Var}(\hat S_y)+S^2 \frac{n_c^2}{n_0^2}\,\mathrm{Var}(f'(\hat S_z))+O(n_c^3)=\frac S 2+n_c^2\,S^2\sum_{k,\ell=2}^{\infty} k\ell\,\chi_k\chi_\ell \bigl(\mu_{k+\ell-2}-\mu_{k-1}\mu_{\ell-1}\bigr)
\label{eq:app_vary_general_unscaled}
\end{equation}

and, therefore, the normalized variance is

\begin{align}
\sigma^2_y(n_c)
&\approx 1+2S\,n_c^2\sum_{k,\ell=2}^{\infty} k\ell\,\chi_k\chi_\ell \bigl(\mu_{k+\ell-2}-\mu_{k-1}\mu_{\ell-1}\bigr).
\label{eq:app_vy_general}
\end{align}

This formula shows that the transverse variance is controlled by the fluctuations of the total shear field $f'(S_z)$, not by a sum of independent contributions from each monomial. In the Gaussian equatorial state, even odd cross terms vanish, so only even even and odd odd sectors remain.

The minimal and maximal variances, i.e., principal variances, are given by the eigenvalues of the normalized covariance matrix in the $y$-$z$ plane \cite{Kit93, Ma10, Li21},

\begin{align}
\Sigma(t)
&=
\begin{pmatrix}
\sigma^2_y(t) & c(t) \\
c(t) & 1
\end{pmatrix}.
\label{eq:app_normalized_cov_matrix}
\end{align}

Hence, the normalized principal variances are

\begin{align}
\sigma^2_{\max,\min}(t)
&= \frac{\sigma^2_y(t)+1}{2} \pm \sqrt{\left(\frac{\sigma^2_y(t)-1}{2}\right)^2+c(t)^2}.
\label{eq:app_varmax_varmin}
\end{align}

Substituting Eqs.~\eqref{eq:app_c_general} and \eqref{eq:app_vy_general} gives the corresponding principal variances to leading order. In particular, when the covariance is linear in $t$ and the excess variance is quadratic in $t$, the principal variances split linearly at short times, as in one axis twisting \cite{Kit93}.

For the special case of a single monomial,

\begin{align}
H
&= \hbar n_0 \chi_k \hat S_z^k,
\label{eq:app_monomial_ham}
\end{align}

one has $f'(S_z)=\hbar n_0 k\chi_k S_z^{k-1}$, so the general formulas reduce to

\begin{align}
c(t)
&\approx -2k\chi_k n_c\,\mu_k,
\label{eq:app_c_monomial}
\end{align}

\begin{align}
\sigma^2_y(t)
&\approx 1+2k^2\chi_k^2 S n_c^2\bigl[\mu_{2k-2}-\mu_{k-1}^2\bigr].
\label{eq:app_vy_monomial}
\end{align}

Using the Gaussian moments gives
\begin{align}
\sigma_y^2(n_c)
&\approx
1+4k^2\left(\frac{S}{2}\right)^k(\chi_k n_c)^2
\begin{cases}
(2k-3)!!, & k \text{ even},\\[1mm]
(2k-3)!!-(k-2)!!^2, & k \text{ odd}.
\end{cases}
\label{eq:app_vy_monomial_gaussian}
\end{align}
The corresponding normalized principal variances follow from substitution into Eq.~\eqref{eq:app_varmax_varmin}.

The range of validity is the initial shearing regime. The derivation relies on a short-time Heisenberg expansion, replacement of $S_x$ by its coherent-state mean, and Gaussian control of the $S_z$ moments \cite{Opa14b, Per19}. For fixed, non-scaled couplings, the relevant small parameter is the accumulated shear across the initial packet, not bare time alone. Once that shear becomes in the order of unity (i.e., the higher cumulants and non-Gaussian distortions become important. This breakdown is familiar already in one-axis twisting \cite{Kit93, Ma10} and is even more pronounced for higher powers, which tend to generate stronger nonlinear deformations of the initially Gaussian packet \cite{uria2023emergence}.

\section{$\chi_k$ limits in two distinct regimes}

The Chebyshev polynomials of the second kind satisfy
\begin{equation}
|U_{k-1}(x)| < k, \quad \text{for } |x|<1, \qquad |U_{k-1}(\pm1)| = k. \label{eq:SM_ChebyshevBounds}
\end{equation}
Since $\mathcal L_a(x_a)\geq0$ for all real $x_a$, the definition of $\mathcal T_0$ gives the unconditional bound
\begin{equation}
\mathcal T_0^{-1} = \Big(1+\tfrac{N\eta}{2}\mathcal L_a(x_a)\Big)^2+\Big(x_c+\tfrac{N\eta}{2}\mathcal L_d(x_a)\Big)^2 \geq \Big(1+\tfrac{N\eta}{2}\mathcal L_a(x_a)\Big)^2 \geq 1, \nonumber
\end{equation}
so that $0<\mathcal T_0\leq1$ for all $\eta$, $N$, $x_a$, and $x_c$, independently of $\mathcal D$. Combining this with~\eqref{eq:SM_ChebyshevBounds} in the closed-form expression for $\chi_k$ yields the rigorous bound
\begin{equation}
|\chi_k| \;\leq\; \frac{k\,\eta^k}{2}\,|\mathcal L_d(x_a)|\,\mathcal L_a(x_a)^{\frac{k-1}{2}},
\label{eq:SM_chi_bound}
\end{equation}
valid for all parameters, with no dependence on $\mathcal D$ or on the asymptotic behavior of $\mathcal T_0$.

We now specialize this bound to the two operating regimes discussed in the main text. In the dispersive cavity regime, $\Delta_{ac}\gg\sqrt{\kappa\Gamma N\eta}$, the atom-cavity detuning satisfies $x_a\gg1$, and the Lorentzian responses reduce to $\mathcal L_a(x_a)\approx x_a^{-2}$, $\mathcal L_d(x_a)\approx -x_a^{-1}$. Substituting into~\eqref{eq:SM_chi_bound} gives
\begin{equation}
|\chi_k| \;\lesssim\; \frac{k}{2}\,\frac{\eta^k}{x_a^{\,k}},
\label{eq:chidispersive}
\end{equation}
independently of $N$ and of the relative size of $\kappa/\Gamma$, since the bound $\mathcal T_0\leq1$ used above already holds unconditionally.

The other regime is realized near the atom--cavity resonance, $\Delta_{ac}\approx0$, with the drive additionally detuned from the dressed cavity resonance by $\delta_R$ much larger than the resonance width. In this case $|x|\approx1$, so the polynomial saturates its bound in Eq.~\eqref{eq:SM_ChebyshevBounds}, and $\mathcal T_0$ approaches its asymptotic value $\mathcal T_0\approx x_c^{-2}$ rather than merely satisfying $\mathcal T_0\leq1$. Using this refined estimate for $\mathcal T_0$ in place of the bound in~\eqref{eq:SM_chi_bound} gives the limiting scaling
\begin{equation}
|\chi_k| \;\approx\; \frac{k}{2}\,\frac{\eta^k}{x_a^{\,k}\,x_c^{\,k-1}},
\label{eq:chinearres}
\end{equation}
which carries the same overall prefactor as Eq.~\eqref{eq:chidispersive}, with the additional suppression $x_c^{-(k-1)}$ arising from the large drive detuning.

\section{Existing Platforms results}
Experimental cavity QED platforms relevant to cavity-mediated collective-spin dynamics. Rates are given as ordinary frequencies. Values marked as inferred are obtained by mapping the reported experimental configuration onto the notation used here.
\begin{table}[h!t]
\centering
\begin{tabular}{lccccc}
\hline\hline
Experiment &
$N$, $\eta$ &
\makecell[l]{ Atom/cavity \\parameters} &
Drive &
$R_{3}/R_{2}$ &
$R_{4}/R_{2}$ \\
\hline
\makecell{Leroux et al.~\cite{Ler09}\\Dispersive} &
\makecell[l]{$N \approx 3.2\times10^{4}$ \\ $\eta=0.139(5)$} &
\makecell[l]{$\Delta_{ac}=3.29\,\mathrm{GHz}$ \\ $\kappa=1.01(3)\,\mathrm{MHz}$ \\ $\Gamma=6.065\,\mathrm{MHz}$} &
$2\delta_{R}/\kappa = 1$ &
$1.09\times10^{-4}$ &
$2.56\times10^{-8}$ \\[2mm]\hline

\makecell{Hosten et al.~\cite{hos16}\\Dispersive} &
\makecell[l]{$N=5\times10^{5}$ \\ $\eta\approx 3.2$} &
\makecell[l]{$\Delta_{ac}\approx 3.42\,\mathrm{GHz}$ \\ $\kappa=10.4\,\mathrm{kHz}$ \\ $\Gamma=6.065\,\mathrm{MHz}$} &
$2\delta_{R}/\kappa = 8$ &
$2.0\times10^{-5}$ &
$1.0\times10^{-9}$ \\[2mm]\hline

\makecell{Davis et al.~\cite{Dav20} \\Dispersive} &
\makecell[l]{$N\sim10^{5}$ \\ $\eta=5.1$} &
\makecell[l]{$\Delta_{ac}=-10\,\mathrm{GHz}$ \\ $\kappa=0.2\,\mathrm{MHz}$ \\ $\Gamma=6\,\mathrm{MHz}$} &
$2\delta_{R}/\kappa = 11$ &
$3.46\times10^{-5}$ &
$6.74\times10^{-8}$ \\[2mm]\hline

\makecell{Greve et al.~\cite{Gre22} \\Dispersive} &
\makecell[l]{$N\approx 900$ \\ $\eta\approx 2.2$} &
\makecell[l]{$\Delta_{ac}=350\,\mathrm{MHz}$ \\ $\kappa=56(3)\,\mathrm{kHz}$ \\ $\Gamma=6\,\mathrm{MHz}$} &
$2\delta_{R}/\kappa = 2.7$ &
$6.0\times10^{-3}$ &
$7.6\times10^{-5}$ \\[2mm]\hline

\makecell{Braverman et al.~\cite{Bra19}\\ Resonant} &
\makecell[l]{$N=1000$ \\ $\eta=1.8$} &
\makecell[l]{$\Delta_{ac}\approx 0$ \\ $\kappa=0.500\,\mathrm{MHz}$ \\ $\Gamma=0.184\,\mathrm{MHz}$} &
\makecell[l]{ $\delta_{R}=2.7\,\mathrm{MHz}$ \\$x_{c}\approx 17$ \\ $x_{a}\approx 79$} &
$1.41\times10^{-2}$ &
$4.63\times10^{-4}$ \\[2mm]\hline

\makecell{Colombo et al.~\cite{Col21}\\Resonant} &
\makecell[l]{$N=100$--$400$ \\ $\eta=7.7$} &
\makecell[l]{$\Delta_{ac}=0$ \\ $\kappa=0.530\,\mathrm{MHz}$ \\ $\Gamma=0.184\,\mathrm{MHz}$} &
\makecell[l]{$\chi_{c}\approx 30$ \\ $\chi_{a}\approx 87$ \\ $\delta_{R}=3.5\,\mathrm{MHz}$} &
$1.03\times10^{-2}$ &
$2.50\times10^{-4}$ \\
\hline\hline
\end{tabular}
\caption{Comparison of experimental cavity QED parameters and the relative strength of higher-order nonlinearities.}
\end{table}

\end{document}